# On the acceptance, commissioning, and quality assurance of electron FLASH units

Allison Palmiero[1]*, Kevin Liu[2,3]*, Julie Colnot[4]*, Nitish Chopra[2], Denae Neill[2], Luke Connell[2,3], Brett Velasquez[2], Albert C. Koong[5], Steven H. Lin[5], Peter Balter[2], Ramesh Tailor[2], Charlotte Robert[4], Jean-François Germond[6], Patrik Gonçalves Jorge[6], Reiner Geyer[6], Sam Beddar[2,3#,] Raphael Moeckli[6#], and Emil Schüler[2,3#]

[1]Department of Radiation Oncology, James Cancer Hospital and Solove Research Institute, The Ohio State University, Columbus, Ohio, USA.

[2]Division of Radiation Oncology, Department of Radiation Physics, The University of Texas MD Anderson Cancer Center, Houston, Texas, USA

[3]Graduate School of Biomedical Sciences, The University of Texas, Houston, Texas, USA

[4]INSERM U1030, Gustave Roussy, Université Paris-Saclay, Villejuif, France

[5]Division of Radiation Oncology, Department of Radiation Oncology, The University of Texas MD Anderson Cancer Center, Houston, Texas, USA

[6]Institute of Radiation Physics, Lausanne University Hospital and Lausanne University, Rue du Grand-Pré-1, Lausanne CH-1007, Switzerland

* Allison Palmiero, Kevin Liu, and Julie Colnot all contributed equally to this work.

# Co-senior/Co-corresponding authors:

| | |
|---|---|
| Emil Schüler, PhD, and Sam Beddar, PhD.<br>Department of Radiation Physics<br>Division of Radiation Oncology<br>The University of Texas MD Anderson Cancer Center<br>1515 Holcombe Blvd<br>Houston, TX 77030, USA<br>Email: eschueler@mdanderson.org,<br>asbeddar@mdanderson.org | Raphael Moeckli, PhD.<br>Institute of Radiation Physics<br>Lausanne University Hospital<br>Lausanne University<br>Rue du Grand- Pré-1<br>Lausanne CH-1007, Switzerland<br>Email: raphael.moeckli@chuv.ch |

**Abstract**

**Background & Purpose**: FLASH or ultra-high dose rate (UHDR) radiation therapy (RT) has gained attention in recent years for its ability to spare normal tissues relative to conventional dose rate (CDR) RT in various preclinical trials. However, clinical implementation of this promising treatment option has been limited because of the lack of availability of accelerators capable of delivering UHDR RT. Commercial options are finally reaching the market that produce electron beams with average dose rates of up to 1000 Gy/s. We established a framework for the acceptance, commissioning, and periodic quality assurance (QA) of electron FLASH units and present an example of commissioning.

**Methods**: A protocol for acceptance, commissioning, and QA of UHDR linear accelerators was established by combining and adapting standards and professional recommendations for standard linear accelerators based on the experience with UHDR at four clinical centers that use different UHDR devices. Non-standard dosimetric beam parameters considered included pulse width, pulse repetition frequency, dose per pulse, and instantaneous dose rate, together with recommendations on how to acquire these measurements.

**Results:** The 6- and 9-MeV beams of an UHDR electron device were commissioned by using this developed protocol. Measurements were acquired with a combination of ion chambers, beam current transformers (BCTs), and dose-rate–independent passive dosimeters. The unit was calibrated according to the concept of redundant dosimetry using a reference setup.

**Conclusions**: This study provides detailed recommendations for the acceptance testing, commissioning, and routine QA of low-energy electron UHDR linear accelerators. The proposed framework is not limited to any specific unit, making it applicable to all existing eFLASH units in the market. Through practical insights and theoretical discourse, this document establishes a benchmark for the commissioning of UHDR devices for clinical use.

# 1. INTRODUCTION

FLASH radiotherapy (RT) has gained attention in recent years for its promise in delivering radiation doses to the treatment volume in less than a second while providing significant normal tissue sparing compared with conventional dose rate (CDR) RT without compromising the tumoricidal effect, as evidenced by numerous preclinical studies.[1-5] This "FLASH effect" is achieved with ultra-high dose rate (UHDR) beams that deliver a mean dose rate of at least 40 Gy/s for a total duration of less than 200 ms.[5] The first studies investigating the FLASH effect used either prototype linear accelerators or existing clinical linear accelerators converted to produce UHDR by increasing the beam current (electron gun), increasing the radiofrequency power (klystron or magnetron), and removing attenuators in the linac head such as flattening filters and collimators to increase the beam output.[6-8] Alternative means of achieving UHDR beams are available in dedicated experimental systems such as the IntraOp Mobetron unit[9-11] (IntraOp, Sunnyvale, CA, USA), the Oriatron eRT6[12] (PMB ALCEN, Peynier, France), the FLASHKNiFE[13] (THERYQ, Peynier, France) and the ElectronFLASH linac[14] (SIT–Sordina IORT Technologies, Vicenza, Italy). The expansion of commercial systems producing UHDR beams, with the end goal being the clinical translation of FLASH RT, underscores the need for guidelines for the commissioning of FLASH-capable devices to demonstrate their reliability in dose delivery and output. Because FLASH RT is a relatively new field, literature is sparse on the elements of acceptance testing and commissioning of electron FLASH (eFLASH) units.[9]

This report summarized the joint experience of four clinical centers to provide a comprehensive framework that parallels the established literature on clinical electron beam dosimetry, such as AAPM Task Group 25,[15] quality assurance (QA) testing according to guidelines from AAPM Task Group 142[16] and machine acceptance, commissioning, and QA from Task Group 72,[17] and recommendations from the IEC 60976[18] and 60977[19] standards on the functional performance characteristics of medical electron accelerators. This report provides guidance for acceptance testing, commissioning, and implementing quality controls for eFLASH units and provides an example outlining our framework for commissioning by using a Mobetron unit. These guidelines can be taken as a starting point for developing unit-specific protocols for commissioning and calibration of other FLASH-capable units.

# 2. METHODS

## 2.1. Guidelines for acceptance, commissioning, and periodic QA of an electron FLASH unit

### 2.1.1. Radiation protection

Radiation protection is a key aspect of the implementation of any eFLASH unit, as these units are capable of delivering substantial doses if irradiation time is not strictly controlled. Therefore, the conceptual design of the machine and the acceptance, commissioning, and QA measurements should be developed with radiation protection limits in mind. Moreover, if the system is intended for use in surgical operating rooms or in a low-shielding area, alternate rooms must be considered for performing commissioning and QA. In any case, for UHDR RT, dosimetry as performed for CDR RT is not possible, standard clinical tools (such as scanning water tank) will not be suitable, and specific dosimeters will have to be used instead. Moreover, radiation safety considerations argue for limiting the beam time for QA as much as possible, and for a specific protocol to be set. Consideration also needs to be taken for detectors conventionally used for radiation protection measurements and their appropriate use in UHDR.

The authors recommend the following minimum requirements regarding radiation protection according to their state and national regulations:

1. The weekly workload should be defined according to the limits set by the local authorities.
2. A complete radiation protection report should be submitted to the authorities before any beam is used clinically. As an example, if the shielding is insufficient, an organization plan should be provided that prevents any person from being in the sector during the irradiation.
3. A radiation survey must be carried out as soon as possible when a beam is available.
    a. The survey will be done at the highest energy available, for the worst-case scenario (maximum dose rates), and in a room configuration resembling that of future use (in particular regarding accelerator location and beam angle). Measurements will be obtained in all surrounding rooms and at every transit point (hallway, control room) or weakness in shielding (door, holes for cable entry, etc.).
    b. The potential for neutron activation of any linac components when energies higher than 10 MeV are involved must be considered.
    c. Means of monitoring the workload with passive dosimetry may be required by authorities. Active dosimetry of all workers may also be requested by authorities.

2.1.2. General guidelines on UHDR detectors and beam reference dosimetry

As long as no primary reference for UHDR beams has been established, reference dosimetry should be performed by redundancy, i.e., by using multiple dosimeters, preferably with different physics concepts for dose measurement, and by checking the compatibility of the results in terms of uncertainty. In the literature, combinations of alanine, thermoluminescent dosimeters (TLDs), optically stimulated luminescent dosimeters (OSLDs), Gafchromic films,[20-23] and active detectors[24] have proven suitable for dosimetry of UHDR beams and have been used for redundant dosimetry.[25] We recommend that redundant dosimetry include three dosimetric systems, each having different detection principles. Once a track record has been established within the beam parameter space of the specific UHDR unit, the number of dosimetric systems may be scaled down. Traditional ion chambers, although of limited use for reference dosimetry, can be used for beam monitoring and QA at extended source-to-surface distances (SSDs) or in the bremsstrahlung tail of the electron percent depth dose (PDD). Likewise, beam current transformers (BCTs) can be used for beam monitoring, with the added benefit of high temporal resolution that allows beam monitoring and QA of individual pulses.[10,11] Because the usual reference conditions may not be reached, the reference dosimetry should be done under "local" reference conditions (as determined by the user) that may differ from one device to the other. The "local" reference conditions should be fully described.

2.1.3. Acceptance testing

The acceptance testing protocol should be specific to the unit and the vendor and may undergo modifications over time. However, the following minimum set of items should be included in the acceptance testing procedure:

1. Interlocks, safety features, and mechanical testing
2. Beam characteristics tuning (if not previously performed at factory)
3. Beam characteristics validation
4. Beam monitoring validation
5. Console functionalities check
6. Docking system tests if applicable
7. Options and accessories functionalities evaluation

*2.1.3.1. Interlocks, safety features, and mechanical testing*

All interlocks and safety features should be tested as part of the manufacturer's acceptance testing procedures. The recommended tests are described in Table 1.

**TABLE 1**. Interlocks and mechanical tests to be used during acceptance testing.

| Tests | Description | Tolerances |
|---|---|---|
| Mechanical inspection | Verification of the movement range, speed, accuracy of the gantry, of the whole unit, of the control unit if applicable and of the beam stopper if indicated<br>Verification of the physical sizes of all applicators | According to manufacturer specifications and tolerances |
| Control console | Verification of the normal operation of each control console function (e.g., interlocks, beam configuration, beam generation, beam monitoring) | Functional |

| Docking system | Verification of the normal function of the docking system (soft or hard-docking system if applicable, automatic applicator recognition) | Functional |
|---|---|---|
| Options and accessories | Verification of normal function (laser, source-to-surface distance indicator, light field) | Functional |
| Safety features | Examination of all safety features (emergency off, beam-on light, door safety, and audible warning sounds) | Functional |

*2.1.3.2. Beam characteristics tuning*

Beam tuning should be done by the manufacturer for each available beam energy and mode, for CDR and UHDR (if both are available). In general, matching both the CDR and UHDR PDD and profiles would be useful. Beam tuning includes adjustments of the beam energy, output rate, and flatness and symmetry of the reference applicator used for output calibration.

*2.1.3.3. Beam characteristics validation*

In addition to beam characteristics for standard medical linear accelerators, the following beam characteristics are recommended for UHDR units:

1. Dose rate under reference conditions: dose per pulse (DPP) and average dose rate (ADR)
2. DPP repeatability and reproducibility
3. DPP proportionality as a function of the number of pulses
4. DPP proportionality as a function of pulse width (PW) and pulse repetition frequency (PRF)
5. Dosimetry interlock (maximum number of pulses allowed)

The full list of recommended tests is shown in Table 2 for UHDR beams, along with recommended tolerances. The specifications and tolerances of the manufacturer must be used if they differ from these values. For UHDR measurements, the lowest possible number of pulses should be used for reasons of radiation protection.

*2.1.3.4. Console, docking, and accessory functionality*

Functionality of the console located outside of the vault must be validated for each type of control. This includes switching between energies (FLASH and CDR), PWs, PRFs, and verifying that the number of pulses or monitor units are delivered appropriately. Mobile systems with docking functionality require acceptance testing of the docking system. For instance, for the Mobetron, this consists of rotation, tilt, and translational shifts of the gantry correlating to the LED display as a guide. The beam characteristics noted previously should also be characterized under imperfect docking conditions. Acceptance testing of all accessories supplied should include individual examination for manufacturer specifications, operation controls, and interlocking capabilities.

**TABLE 2**. UHDR beam characteristics to be tested during acceptance.

| Tests | Description | Tolerances | Comments and recommendations |
|---|---|---|---|
| Reproducibility | 10 consecutive irradiations<br>All energies<br>Reference conditions | 0.5% | Recommended number of pulses: 3 pulses |
| Proportionality of the dose monitoring system | Dose measurement over a range of pulse numbers using the reference pulse repetition frequency (PRF) and reference pulse width (PW)<br>All nominal energies<br>Reference applicator<br>Evaluation of the discrepancies to the linear fit | 2% | Recommended pulse range: 1–30 pulses |
| Independence of output and dose monitoring system with PRF | Dose measurement over a range of PRFs<br>All nominal energies<br>Reference applicator | 2% | Recommended PRF range: 5 PRFs including min and max values<br>Suggested number of pulses: 3 pulses + 1 measurements with a high number of pulses to test potential frequency change with heating |
| Proportionality with PW | Dose measurement over a range of PWs<br>All nominal energies<br>Reference applicator<br>Evaluation of the discrepancies to the linear fit | 2% | Recommended PW range: 5 PWs including min and max<br>Suggested number of pulses: 3 pulses |
| Output stability with beam angle | 4 angular positions including extreme angles<br>5 measurements/configuration<br>Maximum and minimum nominal energies<br>Reference applicator | 3% | Recommended number of pulses: 3 pulses |
| Percent depth dose | All nominal energies<br>Two applicators | Depth of maximum dose: minimum 0.1 cm<br>Ratio of the practical range and R80: max 1.6<br>Maximum discrepancy between measured value and specification of penetrative quality: 3% or 2 mm | Reference applicator and one selected applicator<br>Minimal recommended sampling depth interval: 5 mm<br>Recommended number of pulses: 3 pulses/depth |
| Stability of beam quality with beam angle | Measurements at two depths: depth of dose maximum and depth of 80% of maximum dose<br>One nominal energy<br>4 angular positions including extreme angles<br>Reference applicator | 2 mm or 1% | Recommended number of pulses: 3 pulses/irradiation |
| Surface dose | No additional measurements (measured in percent depth dose test) | Surface relative dose: max 100% | |
| Flatness/ symmetry | Measurements at 3 depths simultaneously<br>All nominal energies<br>Two applicators | Max distance between 80% isodose and geometrical field projection at R90: 15 mm<br>Max distance between 90% isodose and geometrical field projection at reference depth: 10 mm | Reference applicator or the largest if the reference is chosen by the user<br>Depths IEC 60976: Surface (0.5 mm), reference depth R90<br>Recommended number of pulses: 3 pulses/irradiation |
| | | Symmetry: max ratio 105% | |
| Deviation of dose distribution with angular positions | Measurements at 3 depths simultaneously 4 angular positions All nominal energies Reference applicator | 3% | Depths IEC 60976: Surface (0.5 mm), reference depth R90 Recommended number of pulses: 3 pulses/irradiation |

2.1.4. Commissioning

The commissioning phase is the most demanding in terms of dose and UHDR configurations, prompting the need for a different approach for obtaining these measurements. The tests to be done during commissioning as described here are based on the recommendations provided by AAPM Task Group 142 and Table III of the Task Group 72 report[17] and adapted for UHDR requirements. In addition to these basic measurements, it is highly recommended to follow up on the behavior of the machine's stability in terms of output and energy.[26] That follow-up should comprise as many daily checks throughout the commissioning process. Notably, the commissioning should be performed for any modality (CDR and UHDR) and energy that are expected to be used, and should correspond to the modalities and energies validated during acceptance.

The general tests below are recommended along with the tests detailed in Table 3:
1. Daily check to determine the long-term stability of output and energy of the UHDR beams
2. Repeatability between successive measurements
3. PDD and profiles of all possible beam set-up configurations (e.g., energy, size, air gap, collimator), including:
    a. Mean energy evaluation at the surface (calculation based on R50 determination: $E_0 = 2.33 * R_{50}$)
    b. PDD for small and large PWs: with checks of distal depth at 90% of the maximum dose (R90)
    c. PDDs for low and high PRFs
    d. Profiles at a minimum of two depths: the depth of maximum dose ($d_{max}$) and at the depth of 30% of the maximum dose (R30) – considered as a clinically relevant low dose
    e. Output factors

**TABLE 3**. Tests to be done during commissioning of an electron FLASH beam.

| Tests | Suggested dosimeters | Description | Comments and recommendations |
|---|---|---|---|
| Output stability | Active detector suitable for UHDR | 5 consecutive irradiations All nominal energies | Recommended number of pulses: 3 pulses Reference data for daily QA setup |

|  | Films<br>Advanced Markus chamber (large source-to-surface distance only) | Reference applicator<br>Conditions of reference for daily QA | To be repeated each day of the commissioning with long-term stability to be established during routine QA |
|---|---|---|---|
| Energy stability | Active detector suitable for UHDR<br>Films | Energy indicator: ratio of measurements at two depths<br>3 consecutive irradiations at each depth<br>All nominal energies<br>Reference applicator<br>Conditions of reference for daily QA | Recommended number of pulses: 3 pulses<br>Reference data for daily QA setup<br>To be repeated each day of the commissioning with long-term stability to be established during routine QA<br>Recommended depths: reference depth and reference depth*2 |
| Weekly cross profile and PDD follow-up | Films | One cross profile at the reference depth and PDD<br>Reference applicator<br>One nominal energy | One cross profile at the reference depth and one PDD per week during the commissioning<br>Reference data for monthly QA setup |
| PDD | Films or active detector suitable for UHDR | All nominal energies<br>All applicators used for treatments | Redundancy of dosimeters is not mandatory for relative dose measurements if choice of dosimeter has been previously characterized in the beam parameter settings<br>Minimum recommended sampling: 2 mm |
| Profiles | Films | All nominal energies<br>3 depths measured simultaneously<br>All applicators used for treatments | Recommended depths: Surface (0.5 mm), reference depth, R50 |
| Reference dose | 3 independent dosimeters (e.g., alanine, film, TLD, active detector suitable for UHDR) | All nominal energies<br>3 measurements/configuration | The three dosimeters should be irradiated simultaneously whenever possible.<br>Recommendation: 2 PWs, 2 numbers of pulses |
| Output factors | 2 dosimeters (e.g., films, active detector suitable for UHDR) | All regular applicators<br>All nominal energies<br>Different PW and PRF<br>3 measurements/configuration | Measurement at the reference depth |
| Air gap factor | 2 dosimeters (e.g., films, active detector suitable for UHDR) | Two applicators<br>All nominal energies<br>4 gaps<br>3 measurements/gap | Recommended applicators: reference and diameter that will be most commonly used in clinical setup<br>3 measurements/gap |
| Deviation of PDD with UHDR beam parameters | Films | Energy indicator: ratio of measurements at two depths<br>Reference applicator<br>All nominal energies | Recommendation: min, max and median PRF and PW<br>Number of pulses to deliver: 3 different numbers of pulses |

2.1.5. QA

UHDR units are typically less stable than CDR linacs, so the expected tolerances of the machine are to be altered based on the stability level the machine is capable of. The output as well as the energy consistency should be checked daily, as the output may not be as stable as a conventional machine equipped with monitoring ion chambers. If the energy proves to be sufficiently consistent, then the frequency could reasonably be reduced to monthly after proper documentation of its consistency.

Because the docking mechanism could affect the symmetry and flatness of the beam, it should be checked monthly.

In general, a low number of pulses (10 to 20 or even fewer when possible) are used for radiation protection reasons. The recommended tests and their frequencies and tolerances are shown in Tables 4, 5, and 6.

**TABLE 4**. Daily QA checks.

| Tests | Description | Comments and recommendations |
|---|---|---|
| Output and energy stability | Energy indicator: ratio of measurements at two depths<br>Reference applicator<br>Reference UHDR parameters<br>All nominal energies | Recommendation: several measurements to also assess short-term stability (from 5-10 measurements/point at first to 3 with consistency experience)<br>Recommended depths: reference depth and 2*reference depth |
| Interlocks and mechanical | Door interlock, emergency off, collisional interlocks<br>Docking system inspection<br>Mechanical motion (in every degree of freedom), source-to-surface distance indicator if applicable | |

**TABLE 5**. Monthly QA checks.

| Tests | Description | Comments and recommendations |
|---|---|---|
| Output and energy stability | Daily QA<br>Reference applicator<br>All nominal energies | Monthly follow-up with a dose representative of foreseen use |
| Flatness / symmetry in reference condition | Reference applicator<br>All nominal energies | |
| Stability follow-up with different UHDR parameters | Reference applicator<br>All nominal energies | Recommended UHDR parameters: min and max pulse repetition frequency and pulse width, number of pulses: 2 |
| Profile follow-up | One nominal energy<br>Applicator the most used in clinic | |
| Interlocks and mechanical | Daily QA<br>As applicable: light field, centering laser, etc. | |

**TABLE 6**. Annual QA checks.

| Tests | Description | Comments and recommendations |
|---|---|---|
| Output calibration for reference conditions | Reference applicator<br>All nominal energies<br>3 irradiations/configuration | Recommendation: at least two (preferably three) independent types of dosimeters should be used<br>Measurements should be conducted with all dosimeters simultaneously whenever possible |
| Percent depth dose in reference conditions | Reference applicator<br>Energies used | |
| Percent depth dose for selected applicators | 4 applicators<br>Energies used | Applicators used in clinic |
| Cross profiles: flatness / symmetry in reference conditions | Reference applicator<br>Energies used | |
| Cross profiles: flatness / symmetry for selected applicators | 4 applicators<br>Energies used | Applicators used in clinic |
| Output factors for selected applicators and air gap factors | 4 applicators<br>All nominal energies<br>3 irradiations/configuration | Applicators used in clinic<br>Two independent types of dosimeters should be used |
| Output factors for selected PW and PRF | 2 dosimeters<br>(e.g., films, active detector suitable for UHDR) | Reference applicator<br>All nominal energies<br>Different PW and PRF<br>3 measurements/configuration |
| Output constancy with beam orientation | Reference applicator<br>4 angulations<br>All nominal energies<br>3 irradiations/configuration | |
| Percent depth dose constancy with beam orientation | Reference applicator<br>4 angulations<br>All nominal energies | |
| Profiles constancy with beam orientation | Reference applicator<br>4 angulations<br>All nominal energies | |
| Proportionality with pulse width and the number of pulses<br>Output independence with PRF | Reference applicator<br>All nominal energies | Complete verification over the range of pulse widths and pulse repetition frequencies (PRFs)<br>Up to a minimum of 30 pulses |
| Interlocks and mechanical | Monthly QA | |

## 2.2. Practical use of the guidelines: implementing the acceptance, commissioning, and QA process for the Mobetron FLASH unit

The IntraOp FLASH Mobetron was used in this study to demonstrate the process of commissioning an electron FLASH unit according to the protocol outlined in section 2.1. The Mobetron eFLASH machine is a compact and mobile commercial linear accelerator capable of delivering pulsed electron beams at CDR (~10 Gy/min) and UHDR (>40 Gy/s) with energies of 6 and 9 MeV (Figure 1, Table 7). Integrated into the irradiation head are one transmission ion chamber for CDR beam monitoring and control, and two BCTs for redundant UHDR beam monitoring.[11]

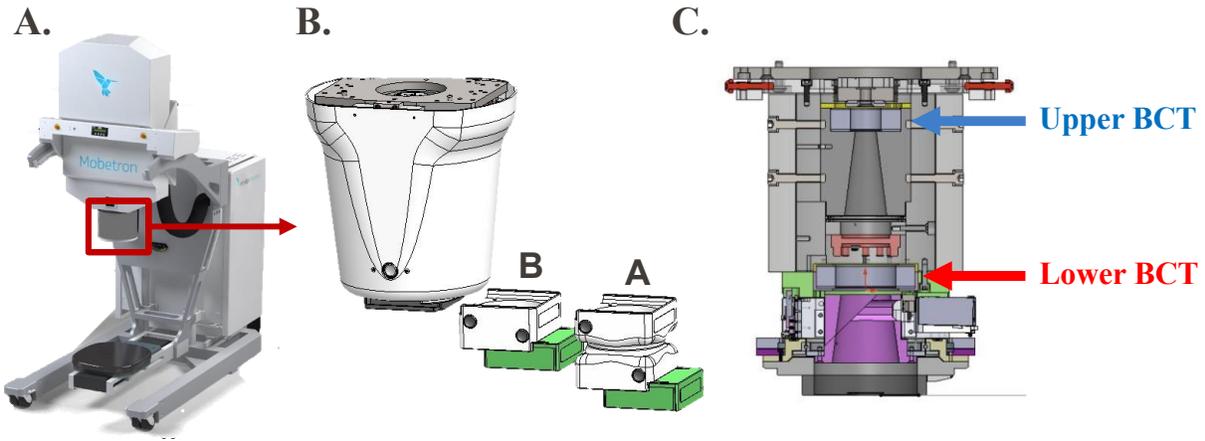

**FIGURE 1.** (A) IntraOp Mobetron unit with (B) exit head, including the A- and B- cones housing their collimator inserts (green), and (C) interior schematic of Mobetron head, with the upper and lower beam current transformers (BCTs) indicated.

**TABLE 7**. Beam parameters of Mobetron FLASH unit

| Parameter | Range |
|---|---|
| Beam energy, MeV | 6 (FLASH) and 9 (CDR and FLASH) |
| Pule width, μs | 0.5–4 |
| Pulse repetition frequency, Hz | 5–120 |
| Gantry tilt | +10°/-30° |
| Gantry rotation | ± 45° |
| Source-to-surface distance for a 5-cm air-gap | 43.7 cm (A-cone) or 38.7 cm (B-cone) |
| Collimator diameters, cm | 2.5–10 |

### 2.2.1. Radiation protection

The Mobetron unit was placed in a pre-existing linac vault originally designed for 18-MV photon beams, and thus no issues with shielding were expected for this unit; this was confirmed in the radiation protection survey. Before the shielding evaluation, the Mobetron unit was pre-tuned in the factory to achieve maximum output using the 9-MeV beam. A radiation survey was done with the anticipated worst-case scenario (maximum output settings). Although patient workload would be considered low, as the device is not currently in clinical use, a high workload was used in barrier calculations, because research throughput would outweigh what would be used for clinical purposes.

2.2.2. Radiation detectors and phantom materials

Electron beam data were collected with a combination of a parallel plate ionization chamber[27] (Advanced Markus (PTW-Freiburg, GmbH, Freiburg, Germany), Gafchromic film,[28] TLDs, OSLDs, and BCTs.[10,11] The local reference conditions for the reference dosimetry were at the depth of maximum dose using a 10-cm diameter insert with a 5-cm air gap between the collimator and the surface of the water/phantom (Table 7). The reference dosimetry was done with film, TLDs, and OSLDs. Gafchromic EBT3 film was also used for relative dose measurements. Percent depth dose curves were generated by placing the film inside a 3D printed adaptation of an in-house water tank, with the film positioned at a 2% angle relative to the central axis of the beam.[29] Beam profiles, radiation field size, and output vs. gantry angle were measured in solid water (Solid Water HE, Sun Nuclear Corporation, Melbourne, FL, USA). An Advanced Markus parallel plate chamber at extended SSD and the BCTs were used for daily constancy measurements and for beam monitoring.

Films were scanned at 24 hours after irradiation on an Epson 10000XL flatbed scanner (Seiko Epson Corporation, Nagano, Japan). Films were scanned at 72 dpi when used for point dose measurements and at 150 dpi for relative dosimetric measurements. Films were analyzed by using the red channel with ImageJ and MATLAB as previously reported.[30] LiF:Mg,TI TLD powder (ThermoFisher, Waltham, Massachusetts, USA) and nanoDot OSLDs (Landauer, Inc., Glenwood, Illinois, USA) were used as redundancy methods for measuring dose delivery under reference conditions. The TLD powder was packaged in 1 cm × 1 cm envelopes and measured 24 hours after irradiation with a Harshaw TLD Model 5500 Reader (ThermoFisher). The signal readout was normalized to the weight of the TLD powder. The OSLDs were measured, at least 10 minutes after irradiation, five times and averaged with the OSLD reader (microSTARii; Landauer, Inc.). The stability of the reader was tested before each use session.[31,32]

2.2.3. Acceptance testing

Both FLASH and CDR modes were acceptance-tested per the company's acceptance parameters (interlocks, safety, mechanical tests, gantry rotational and translational verification, light field vs. radiation field comparison). These acceptance tests were in line with those proposed in sections 2.1.3.1 and 2.1.3.4.

*2.2.3.2. Beam characteristics tuning and validation*

The beam percent depth doses and profiles for the CDR and the UHDR beams were matched for the same beam parameter settings (1.2 µs, 30 Hz) by the vendor. Tests were done as described in Table 3, with the eFLASH mode (6- and 9-MeV) of the Mobetron unit with a PW setting of 1.2 µs and PRF setting of 90 Hz; 45 pulses were delivered for the A-applicator and 25 pulses for the B-applicator unless otherwise specified. All measurements were obtained with 5-cm backscatter.

2.2.4. Commissioning

*2.2.4.1. Short-term output and energy stability*

During the commissioning process, daily output measurements were taken to determine machine performance throughout the commissioning process. This involved measurements obtained with an Advanced Markus chamber at extended SSD (110 cm) at the reference depth for each energy. These measurements were taken with a low number of pulses (10 pulses) to address radiation protection concerns. Furthermore, variation within a single day was evaluated by determining variation in machine output with change of temperature within the linac head. In FLASH mode, these data were also obtained with the BCTs as a secondary evaluation. The ratio of the upper and lower BCT was used to determine energy variation in the beam, as previously described.[11]

*2.2.4.2. Electron beam quality and dose distribution*

The relative dose distribution for different collimator inserts/sizes was investigated as follows. The PDD measurements were obtained by placing EBT3 film as described in section 2.2.2, with an SSD of 43.7 cm (A-cone) or 38.5 cm (B-cone) (Figure 1 and Table 7), and a 5-cm air gap between the exit window and water surface.[29] The diameter of the collimator inserts ranged from 2.5 cm to 10 cm.

PDDs were measured with film in the previous irradiation set-up with the A-applicator and 10-cm insert to investigate the relative dose distribution as a function of PRF (25 pulses delivered with a PRF of 5–120 Hz [with a fixed PW of 1.2 µs]), to investigate the relative dose distribution as a function of PW (PRF of 90 Hz for PWs of 0.5, 1.0, 1.2, 2.0, 3.0, and 4.0 µs with the corresponding number of pulses [60, 30, 30, 20, 15, and 10 pulses]) to roughly match the same dose delivered to film for each PW, and to investigate the relative dose distribution for the 9-MeV beam as a function of accelerator temperature (a temperature monitor was attached to the Mobetron head, and irradiations of 30 pulses were performed at temperatures of 26°–32°C, representing a cold-start and the maximum temperature achieved after heavy usage).

*2.2.4.3. Flatness and symmetry*

EBT3 films were placed in solid water at a depth of 2 cm for 9 MeV and 1.5 cm for 6 MeV with a 5-cm air gap between the buildup and collimator insert. The collimator inserts ranged from 2.5 cm to 10 cm and were inserted into either the A- or B-applicator. Flatness and symmetry were calculated according to the Varian definitions. Varian defines symmetry as the difference between dose at some distance from the central axis relative to that on the central axis and flatness as the ratio of difference between maximum and minimum doses to the addition of those same doses. Both flatness and symmetry are defined within 80% of the FWHM.

*2.2.4.4. Output factors*

Output factors for each collimator size ranging from 2.5 cm to 10 cm for both A- and B- applicators were obtained by measuring the dose with EBT3 films at $d_{max}$ with 5-cm backscatter. All readings were normalized to the 10-cm cone for each applicator. To examine the dependence of output on the rotation angle of the gantry, a set number of pulses was delivered to EBT3 films placed between solid water slabs at the depth of maximum dose in an in-house 3D printed holder that was attached directly to the Mobetron head. For each beam energy, measurements in triplicate were performed at 0°, at the maximum gantry tilt angles, and at the maximum gantry rotation angles (Table 7).

To investigate output repeatability and linearity, an Advanced Markus chamber was placed in solid water at 110 cm SSD at depths of 1.5 cm for the 6-MeV beam and 2 cm for the 9-MeV beam. The integrated BCTs were used as a secondary system to assess output and linearity, and the signal ratio of the ion chamber and the BCTs was recorded for each delivery condition to evaluate consistency between the readout systems. Output linearity was evaluated by delivering in triplicate 1, 2, 5, 10, 20, 50, and 100 pulses. Output repeatability was assessed by delivering three pulses (1.2 µs, 30 Hz) five times for each energy and mode throughout the commissioning process.

2.2.5. Quality Assurance

A QA program was implemented for both FLASH and CDR mode according to the recommendations in Tables 4, 5, and 6.

**3. RESULTS**

**3.1. Radiation protection and acceptance testing**

Survey and leakage measurements were taken, and no dose excess was found around the bunker. The door interlock and docking system were all shown to be functional. Gantry and collimator readouts were determined to be within a degree at mechanical limits. The translational motion at mechanical limits agreed within 1 mm for the three directions. The light/radiation field coincidence were determined to be within 2 mm. The distance check device between both the internal and external laser device were compared and determined to be within a 1-mm tolerance at isocenter and extended SSDs. Both FLASH and CDR mode beams met the company's acceptance parameters.

### 3.2. Beam commissioning

All results presented in this section correspond to the FLASH beams of the Mobetron unit. Output variation day to day across the time frame for commissioning was found to be within 3% as determined through BCTs (Figure 2). When output and energy variation was characterized as a function of the machine temperature within a single day, no correlation was found along with consistent PDDs across all temperatures measured (Figure 3).

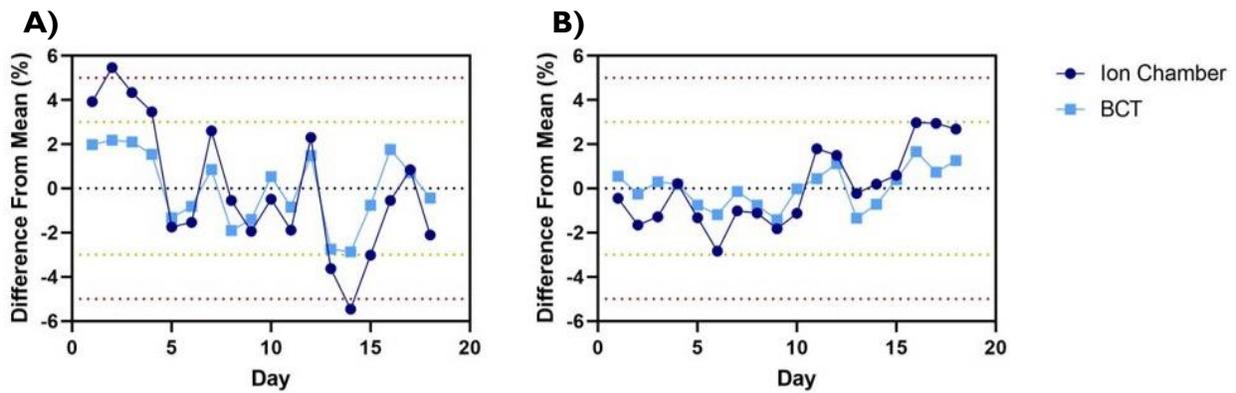

**FIGURE 2**. Short-term stability for the (A) 6-MeV and (B) 9-MeV FLASH beams for both ion chamber and beam current transformers (BCT) measurements. The measurements were obtained daily over a 18-day period. Data are mean ± standard deviation (error bars may be hidden by the measurement points because of their relatively small values.)

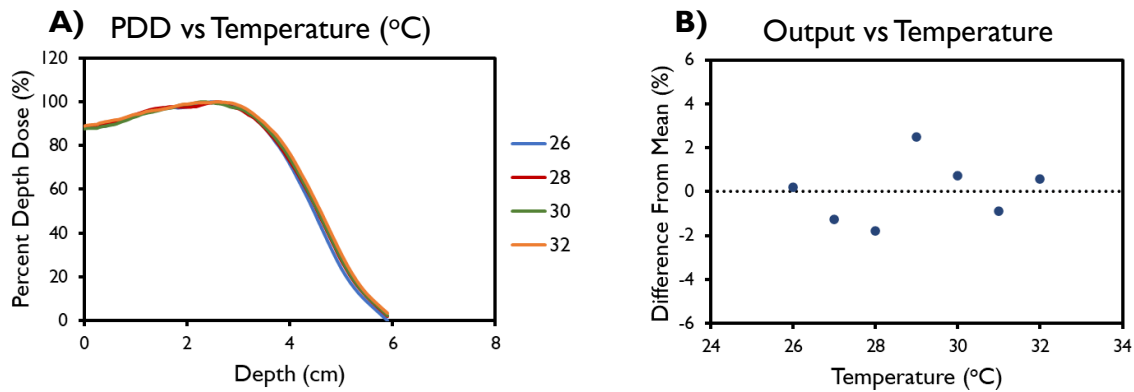

**FIGURE 3.** (A) Percent depth dose (PDD) and (B) variation in output for linac head temperatures between 26°C and 32°C. Data in (B) are mean ± standard deviation (error bars may be hidden by the measurement points because of their relatively small values.)

    During commissioning, PDD analyses were done with varying PRFs, PWs, and field sizes (Figure 4). Decreasing field size was found to correlate with a shallower depths of max dose, an increase in the surface dose, and a reduction of the sharpness of dose fall-off for both tested energies. The 9-MeV beam also had a greater maximum depth shift and decreased sharpness in fall-off compared with the 6-MeV beam for the same cone size. The $R_{50}$ values using the A cone and 10-cm collimator were found to be 3.76 cm for the 9-MeV beam and 2.74 cm for the 6-MeV beam, translating to corresponding $E_0$ values of 8.76 MeV and 6.38 MeV. PDD values remained constant with varying PRF for both energies. However, for both the 6-MeV and 9-MeV beams, the energy of the beam decreased with increasing PW, as evidenced by its shallower $d_{max}$ and shorter $R_{50}$. The dose fall-off was also steeper for higher PWs.

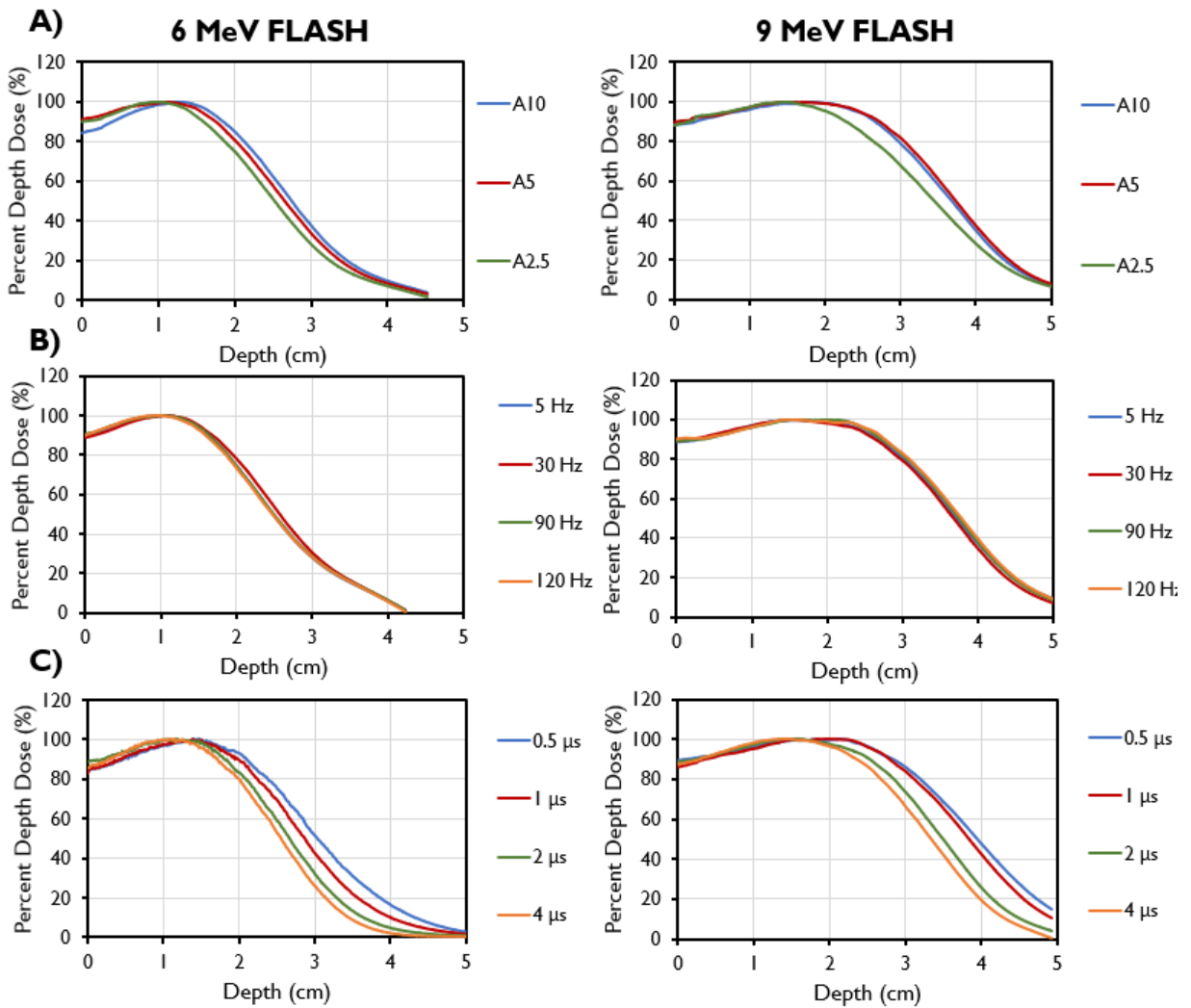

**FIGURE 4.** Percent depth dose curves for 6-MeV beam (left) and 9-MeV beam (right) measured for different collimator sizes using the A-cone (A10 is the A cone with a 10-cm collimator; A5 is the A cone with a 5-cm collimator; A2.5, is the A-cone with a 2.5-cm collimator). (B) PDDs by pulse repetition frequencies (PRFs). (C) PDDs by pulse widths (PWs).

Selected transverse (crossline) profiles for different field sizes (A-cone) and their associated characteristics are presented in Figure 5 and Table 8 (inline data not shown). These data were obtained at reference depths of 1.5 cm for the 6-MeV beam and 2 cm for the 9-MeV beam.

**TABLE 8.** Crossline profiles characteristics measured for both energies of the Mobetron unit.

|  | A-cone, 10 cm collimator | A-cone, 5 cm collimator | A-cone, 2.5 cm collimator |
|---|---|---|---|
| **FWHM, cm** | | | |
| 6 MeV | 10.9 | 5.7 | 2.8 |
| 9 MeV | 10.8 | 5.6 | 2.8 |
| **Crossline symmetry, %** | | | |
| 6 MeV | 0.7 | 2.1 | 4.3 |
| 9 MeV | 1.6 | 1.4 | 1.7 |
| **Crossline flatness, %** | | | |
| 6 MeV | 14.6 | 9.2 | 18.1 |
| 9 MeV | 13.8 | 10.2 | 18.2 |

FWHM, full width half maximum.

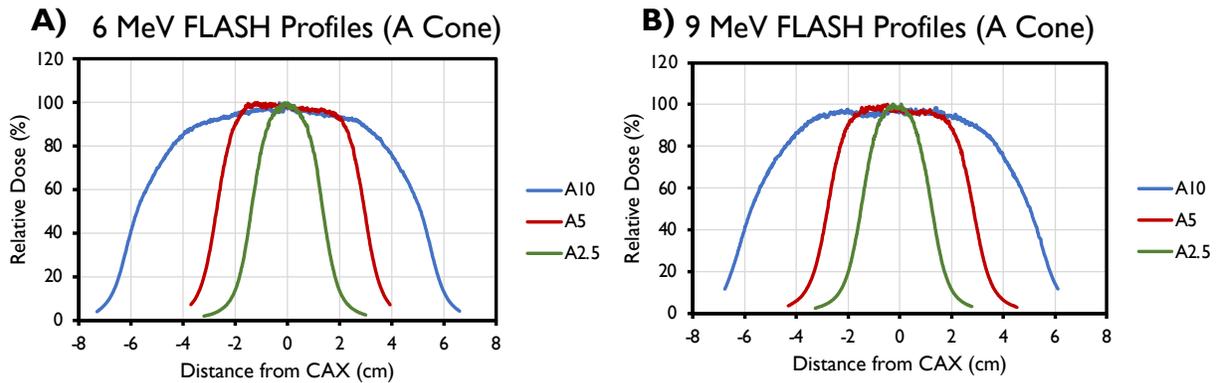

**FIGURE 5**. Central axis crossline beam profiles of the (A) 6-MeV and (B) 9-MeV beams, measured at $d_{max}$ for different field sizes (2.5–10 cm) with the A-applicator.

Other dosimetric parameters measured were linearity with number of pulses, PW, and PRF, and rotational output constancy. Output factors were obtained for every cone size ranging from 2.5 cm to 10 cm and normalized to the 10-cm cone. Output factors for both the A and B cones are presented in Figure 6. The maximum output factors were measured for the 5-cm collimator for the 6-MeV beam and for the 7-cm collimator for the 9-MeV beam, which is consistent with previous reports.[9,33]

Accumulated signal was found to increase linearly ($R^2 = 1$) with an increasing number of pulses (Figure 7). The average ratio of the ion chamber to the BCT was found to be $0.13 \pm 0.001$ (mean ± standard deviation) for the 6-MeV beam and $0.26 \pm 0.003$ (mean ± standard deviation) for the 9-MeV beam. Figure 7 also shows that the signal increases with increasing PW and remains constant with increasing PRF, as expected.

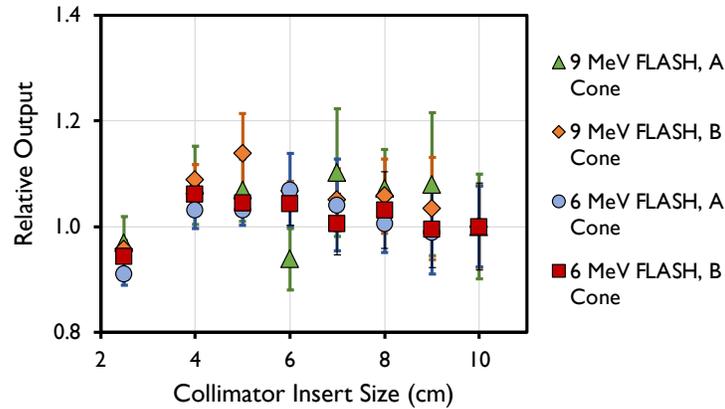

**FIGURE 6.** Output factor of 6-MeV and 9-MeV beams measured at $d_{max}$ for different field sizes (2.5–10 cm) using the A- and B-cone. Data are mean ± standard deviation.

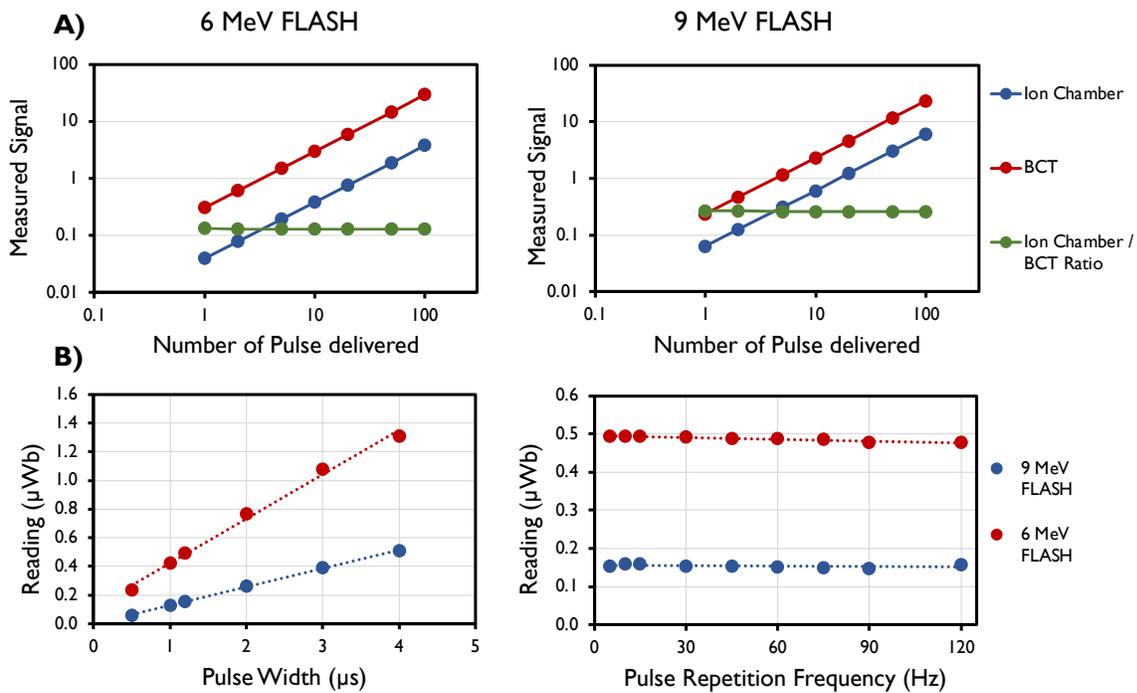

**FIGURE 7.** (A) Linear response measured by ion chamber (Advanced Markus), upper beam current transformers (BCT; μWB), and their ratio for the 6-MeV and 9-MeV eFLASH beams. (B) Linear response with pulse width (PW) and pulse repetition frequency (PRF) measured by upper BCT for the 6-MeV and 9-MeV eFLASH beams. Data are mean ± standard deviation (error bars may be hidden by the measurement points because of their relatively small values.)

### 3.3. Quality assurance program

The data pertaining to output and energy constancy taken as part of the implemented QA program is shown in Figure 8. The data spans a time frame of 2 years, with the exception for the 6 MeV FLASH beam which was decommissioned after 1 year. All beams displayed a high level of stability. Daily output was within 5% of baseline throughout the investigated time period and within 3% of baseline in 93.5%, 93.1%, and 96.8% of days for the 6 MeV FLASH, 9 MeV FLASH, and 9 MeV CONV beams, respectively. The ion chamber data showed a higher level of variability compared to the BCT data, likely due to setup uncertainty. Energy stability throughout the investigated time period was within the recommended 3%/2mm as determined through BCT ratio (as described in Liu et al.[11]) and the ratio of ion chamber measurements in two different depths (data not shown) for the FLASH and CONV beams, respectively.

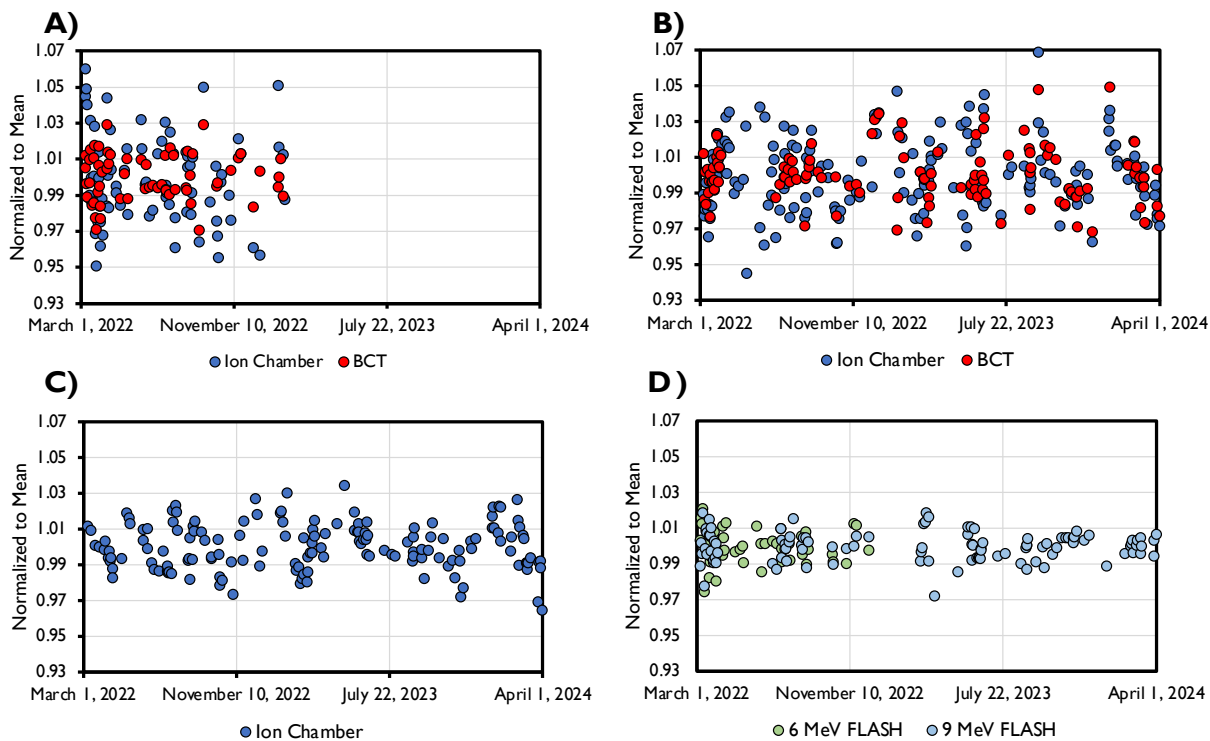

**Figure 8.** Output constancy during a 2-year time frame from the A) 6 MeV FLASH beam, B) 9 MeV FLASH beam, and C) 9 MeV CONV beam. The FLASH beam output constancy data was taken with both ion chamber at extended SSD (110 cm) and through internal upper BCT. The CONV beam constancy data was acquired only with ion chamber. D) Energy constancy, as determined through BCT ratio, for the 6 and 9 MeV FLASH beams.

## 4. DISCUSSION

The recommendations presented here are for dosimetry tests similar to those of Moeckli et. al[9] and included output factors, PDDs, profiles, and linearity on the Mobetron eFLASH unit. The current report expands on that work to include guidelines for acceptance testing, commissioning, and a full QA program, including all dosimetric, mechanical, and safety tests, for eFLASH units intended for clinical use. This report also outlines the recommended dosimeters and how to mitigate potential pitfalls when commissioning an eFLASH machine. The guidelines are based broadly on relevant AAPM documents[15-17,34,35] and IEC recommendations, which should be followed whenever possible. However, several aspects related to eFLASH beams are not covered by currently accepted commissioning and QA protocols for conventional dose rate linear accelerators,[16-19,35] and additional data are required as outlined in this report.

First, eFLASH systems generally allow customization of pulse structure in terms of PW across PRFs, whereas a standard clinical linac does not. For this reason, all commissioning data need to be obtained for each PW and PRF. The options for PW and PRF will vary between different FLASH machines. For this reason, the user needs to determine the proper step sizes in the commissioning process to fully capture the dependence between possible PW and PRF combinations and machine output and energy.

In eFLASH beams, standard ion chambers experience severe ion recombination and are therefore of limited use for dosimetric calibration; however, they are still useful for monitoring short- and long-term stability of beam output and energy. Care is needed, however, that any data acquired with standard ion chambers are measured and validated by using a redundancy approach to ensure that appropriate data are collected. In the commissioning example presented here, we used an Advanced Markus ion chamber because of its well-characterized behavior in eFLASH beams,[27] and we placed it at an extended SSD to avoid severe ion recombination effects. In addition to the Advanced Markus ion chamber, Gafchromic film, TLDs, and OSLDs were used owing to their dose rate–independence and extensive use in FLASH dosimetry.[20,28,30,31] The advantages of using these types of dose rate–independent dosimeters and detectors are their suitability for use at high dose ranges and at UHDR conditions that are pertinent to FLASH RT. The dynamic range of EBT3 film has commonly been reported as less than 10 Gy.[21] However, EBT3 film can be used over a much larger dose range, with some reporting suitable use up to 60 Gy.[27,28,36-39] Similar findings have been presented for OSLDs and TLDs (reviewed in Liu et al[31]). The use of multiple dose rate–independent detectors in a redundancy framework in FLASH beamlines is a necessary tool for accurate dose measurements and calibration, as well as a means of cross-checking and cross-validating the dose delivered. The physical mechanisms of signal generation in Gafchromic film, alanine, TLDs, and OSLDs are different in their own respects, and having redundant tools for measuring dose enables a robust method of performing commissioning and calibration in FLASH beamlines until a reference standard has been established.

BCTs, as a beam monitoring device, have been shown to have a linear response related to dose and DPP and to be independent of mean and instantaneous dose rate; they can be used to monitor the output of eFLASH beams in real time without perturbing the beam.[10,11,40] In this work, BCTs were commissioned for use as a detector option to validate the measurements obtained here. The Mobetron unit, which was used as a practical example in this report, has two BCTs integrated into the head of the unit (Figure 1). This dual-BCT design allows beam output and energy monitoring to be determined while providing redundancy in beam monitoring in real time; it can provide the pulse structure, temporal structure between individual pulses, and beam output by correlating the integrated signal under the pulses to the dose delivered to a dose rate–independent detector at a reference location. Other detectors that can be used for real-time beam monitoring, such as ultra-thin parallel plate ion chambers, diamond detectors, and scintillators, are under development.[24,41-46] Regardless of which real-time beam monitor is chosen, we recommend that users build up their own experience, perform a full characterization, and establish a proven track record in parallel with using redundancy in dosimetric systems with well-established dosimeters. Once this has been established, the number of dosimetric systems can be scaled down.

## 5. CONCLUSIONS

The framework presented here for acceptance testing, commissioning, and QA for eFLASH units represents a consensus framework among four different FLASH RT programs at four clinical centers (two in Europe and two in the United States), with established expertise and long-term experience with various eFLASH units. The proposed framework is not limited to any specific unit but rather provides guidance and practical insight for centers looking to establish a robust framework around eFLASH RT. An example of the practical implementation of these guidelines was described for the Mobetron unit. Thus, this work successfully establishes a robust guidance document for commissioning and QA that can be easily tailored for any eFLASH unit.

**Acknowledgements**

We thank Christine F. Wogan, MS, ELS, of MD Anderson's Division of Radiation Oncology, for editorial contributions to several drafts of this article. Research reported in this publication was supported in part by the National Cancer Institute of the National Institutes of Health under Award Number R01CA266673, by the University Cancer Foundation via the Institutional Research Grant program at MD Anderson Cancer Center, by a grant from MD Anderson's Division of Radiation Oncology, by Cancer Center Support Grant P30 CA016672 from the National Cancer Institute of the National Institutes of Health, to The University of Texas MD Anderson Cancer Center, and by UTHealth Innovation for Cancer Prevention Research Training Program Pre-doctoral Fellowship (Cancer Prevention and Research Institute of Texas grant #RP210042). The content is solely the responsibility of the authors and does not necessarily represent the official views of the National Institutes of Health nor of the Cancer Prevention and Research Institute of Texas.